\documentclass[twocolumn]{aastex631}

\usepackage{amsmath,amssymb}
\usepackage{graphicx}
\usepackage{bm}

\newcommand{\dd}{\mathrm{d}}
\newcommand{\ii}{\mathrm{i}}
\newcommand{\ee}{\mathrm{e}}
\newcommand{\GF}{G_{\mathrm{F}}}
\newcommand{\abs}[1]{\left\lvert#1\right\rvert}

\begin{document}

\title{Inspecting neutrino flavor instabilities during proto-neutron star cooling phase in supernova: \\ I. Spherically symmetric model}

\author[0000-0001-7305-1683]{Masamichi Zaizen}
\email{zaizen@heap.phys.waseda.ac.jp}
\affiliation{Faculty of Science and Engineering, Waseda University, Tokyo 169-8555, Japan}

\author[0000-0001-5031-6829]{Sherwood Richers}
\affiliation{Department of Physics and Astronomy, University of Tennessee, Knoxville, USA}

\author[0000-0002-7205-6367]{Hiroki Nagakura}
\affiliation{National Astronomical Observatory of Japan, 2-21-1 Osawa, Mitaka, Tokyo 181-8588, Japan}

\author[0000-0002-8622-8721]{Hideyuki Suzuki}
\affiliation{Faculty of Science and Technology, Tokyo University of Science, 2641 Yamazaki, Noda, Chiba 278-8510, Japan}

\author[0000-0003-4213-0076]{Chinami Kato}
\affiliation{Faculty of Science and Technology, Tokyo University of Science, 2641 Yamazaki, Noda, Chiba 278-8510, Japan}

\begin{abstract}
In the standard model of core-collapse supernova (CCSN), all neutrinos are assumed to be in pure flavor eigenstates in CCSN cores, but the assumption becomes invalid if neutrino distributions are unstable to flavor conversions.
In this paper, we present a study of the occurrences of two representative neutrino-flavor instabilities, fast- and collisional flavor instabilities, in the cooling phase of proto-neutron star (PNS) from 1- to 50 seconds.
We follow the long-term evolution of a PNS under spherically symmetric and quasi-static approximations, in which the matter profile is determined by solving the Tolman-Oppenheimer-Volkoff equation with neutrino feedback under the treatment of multi-group flux limited diffusion.
For the stability analysis of neutrino flavor conversions, we recompute neutrino distributions using Monte Carlo transport in order to obtain the full angular distribution needed to compute the dispersion relations.
We find no signs of flavor conversions in our models; the physical reason is thoroughly investigated.
We also argue that the negative conclusion in flavor conversions could be changed qualitatively if multi-dimensional effects are included, as similar to cases in the earlier phase of CCSN.
\end{abstract}


\section{Introduction}
Core-collapse supernovae (CCSNe) are dramatic events at the final stage of the stellar evolution for stars with a zero-age main sequence mass of $\gtrsim 10\,M_{\odot}$.
When the central core experiences a core collapse and forms a proto-neutron star (PNS), a huge amount of neutrinos are released, which deleptonize and cools down the CCSN core.
Meanwhile, the shock wave is launched by the core bounce and blows accreting matter off.
Although the shock wave undergoes stagnation or even retraction during its post-bounce evolution, neutrino transport from PNS to the post-shock region and their interactions with matter behind the shock wave via various weak processes (see, e.g., \citealt{Langanke:2003}) can reinvigorate the shock expansion with the aid of multi-dimensional fluid instabilities.
This seems to be a primary driving force of CCSNe for a wide range of progenitor masses (see, e.g., \citealt{Burrows:2020}), which is well known as neutrino-heating mechanism \citep{Janka:2012a,Janka:2017,Mezzacappa:2020,Burrows:2021,Fischer:2024,Yamada:2024}.

In dense neutrino environments such as CCSN cores, refractive effects by neutrino self-interactions, which have been neglected in the standard model of CCSNe, can induce large flavor conversions \citep{Sigl:1993,Duan:2006b,Tamborra:2021,Capozzi:2022,Richers:2022b,Volpe:2024}
Among them is the fast flavor instability (FFI) \citep{Sawyer:2005,Sawyer:2016}, which is triggered by a zero crossing in ELN (electron neutrino-lepton number)-XLN (heavy-lepton neutrino lepton number) angular distribution.
Yet another is the collisional flavor instability (CFI) \citep{Johns:2023a}, which is induced by the disparity in reaction rates between neutrinos and antineutrinos.
Both have attracted significant attention recently due to the potential for instability under the supernova shock.
The appearance of ELN-XLN angular crossing has been explored in state-of-the-art CCSN models \citep{Dasgupta:2017,Abbar:2020,Nagakura:2021b,Akaho:2024a}, and the occurrence of CFI has been also confirmed similarly using linear stability analysis \citep{Liu:2023c,Akaho:2024a}.
These surveys for flavor instabilities imply that flavor conversion ubiquitously occurs in the post-shock region.
Recent studies also suggest that the neutrino radiation field is significantly changed by flavor instabilities, which have an impact on many processes and outcomes in CCSNe such as fluid dynamics \citep{Nagakura:2023,Ehring:2023,Ehring:2023a,Xiong:2023,Xiong:2024,Shalgar:2024}, neutron star kick \citep{Nagakura:2024}, and neutrino signal \citep{Wu:2015,Nagakura:2023d,Ehring:2023,Xiong:2024}.

These recent results of flavor conversions stimulate our interest in the post-explosion phase, as neutrinos continue to play important roles in cooling the remnant PNS and determining nucleosynthesis in the ejecta.
As suggested by earlier studies (see, e.g., \citealt{Takahashi:1994,Qian:1996}), heavy nuclei including (weak) r-process and $\nu$p-process elements could be synthesized in neutrino-driven winds if such winds form (e.g., \citealt{Witt:2021,Wang:2024b,Nevins:2024}).
Nucleosynthesis has, however, a delicate sensitivity to the relative difference between electron-type neutrinos ($\nu_e$) and their anti-partners ($\bar{\nu}_e$) (see, e.g., a recent review of \citealt{Fischer:2020a} and references therein), indicating that it requires accurate treatments of neutrino kinetics that can resolve flavor-dependent features.
This suggests that the occurrence of flavor conversions, if any, could be one of the largest uncertainties in these models.
PNS convection is expected to be present and generate flavor instabilities in the PNS (e.g., \citealt{Glas:2020,Abbar:2020,DelfanAzari:2020}).
Although parameterized models of flavor transformation during PNS cooling cause the amount of ejected mass to increase and more proton-rich conditions, the net effects of flavor instability in self-consistent models are unknown.
This study aims to build on these works by determining whether the conditions for the FFI arise in realistic neutrino distributions during the PNS cooling phase.

In this paper, we present the first detailed study of the occurrence of FFI and CFI in the PNS cooling phase.
We determine the PNS structure and its long-term evolution by our PNS cooling scheme, in which we employ a quasi-static approach and diffusion approximation for neutrino transport under spherical symmetry.
For stability analysis of flavor instabilities, multi-angle neutrino transport simulations are carried out for some selected time snapshots.
We then survey FFI and CFI in the neutrino data using well-established approximate methods to identify these instabilities.

This paper is organized as follows.
We start with introducing essential methodologies to develop models of PNS structure and its long-term evolution in Sec.\,\ref{sec:PNSmod}.
In Sec.\,\ref{sec:Sedonu}, we describe how to model neutrino radiation field under frozen fluid- and metric backgrounds by using a Monte Carlo calculation of neutrino transport.
Briefly summarizing the linear stability analysis of flavor conversion in Sec.\,\ref{sec:Stab}, we show the results in Sec.\,\ref{sec:result}.
Finally, in Sec.\,\ref{sec:conclusion}, we conclude the present study and describe the need for further study towards drawing more robust conclusions.

\section{PNS modeling}\label{sec:PNSmod}

\begin{figure*}[t]
    \centering
    \includegraphics[width=0.9\linewidth]{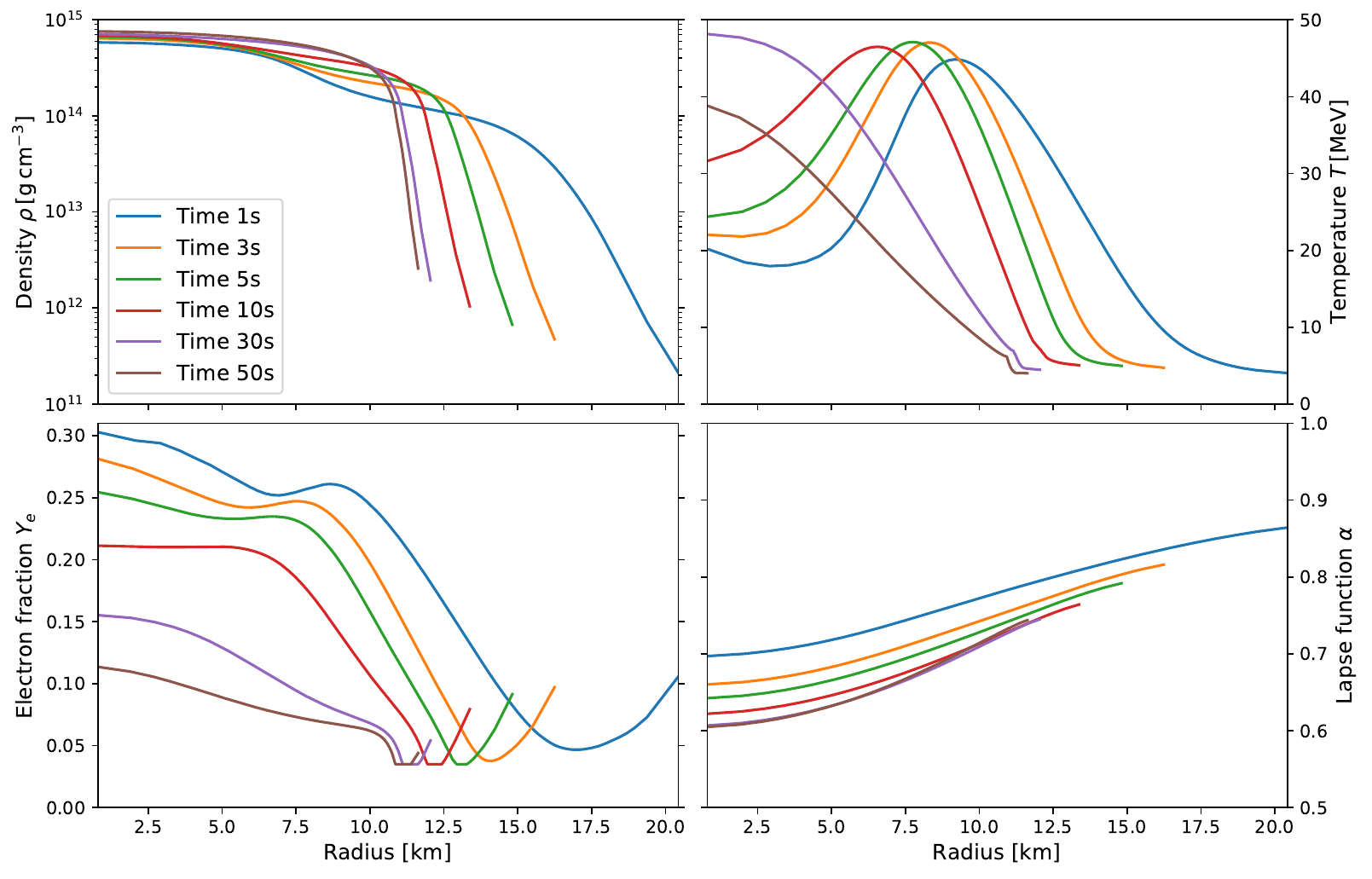}
    \caption{
    Radial profile of baryon mass density, electron fraction, temperature, and lapse function at each time snapshot in our PNS cooling model.
    }
    \label{fig:mat_prof}
\end{figure*}

During cooling stages of the PNS, it is nearly in hydrostatic equilibrium and evolves quasistatically while emitting neutrinos.
In this study, we model the PNS structure at each instant under a condition that the general relativistic spacetime is stationary and spherically symmetric.
The static matter distribution can be, hence, determined by solving Tolman-Oppenheimer-Volkoff (TOV) equation (e.g., \citealt{Burrows:1986,Pons:1999}).

In most of the PNS interior except for the narrow outer envelope, neutrinos interact with the matter on short length scales compared to the system size, and so are effectively trapped.
For this reason, their transport of energy and lepton number in space can be approximated as a diffusion processes.
Since neutrino interaction rates depend sensitively on neutrino energy and flavor, we employ a multi-group (i.e., multi-energy) flux limited diffusion (MGFLD) scheme to determine the neutrino radiation field and its interaction with matter in our PNS cooling model.
We note that a flux limiter is adopted in the region where the diffusion approximation is invalid so as to guarantee that neutrino transport is causal and become accurate in the free-streaming limit.
We refer to Appendix\,\ref{Ap:PNScoolingscheme} for more details of our scheme.

The neutrino-matter interactions in our MGFLD scheme are computed based on the same approach as that in \citealt{Bruenn:1985}, but with some extensions.
We employ the method in \citealt{Suzuki:1993} for the nucleon bremsstrahlung processes which play important roles for production of non-electron type neutrinos during the PNS cooling.
Below we provide a list of weak processes implemented in our code:
\begin{eqnarray}
p + e^{-} &\longleftrightarrow& n + \nu_e, \nonumber \\ 
A + e^{-} &\longleftrightarrow& A' + \nu_e, \nonumber \\ 
n + e^{+} &\longleftrightarrow& p + \bar{\nu}_e, \nonumber \\
\nu + N &\longleftrightarrow& \nu + N, \nonumber \\ 
\nu + A &\longleftrightarrow& \nu + A, \nonumber \\
\nu + e^{\pm} &\longleftrightarrow& \nu + e^{\pm}, \nonumber \\
e^{-} + e^{+} &\longleftrightarrow& \nu + \bar{\nu}, \nonumber \\
N + N' &\longleftrightarrow& N + N' + \nu + \bar{\nu}, \nonumber
\end{eqnarray}
where $p, n, A, N, e$ are protons, neutrons, nuclei, nucleons, and electrons, respectively.
The plasmon decay process into neutrino pairs is also added to the electron-positron pair annihilation processes by scaling the total neutrino emissivity as calculated by \citealt{Kohyama:1986}.

As the initial condition of the PNS, we take a fluid snapshot of a CCSN model simulated by \citealt{Nakazato:2018}.
The CCSN model was developed based on a spherically symmetric general relativistic neutrino-radiation hydrodynamic simulation of a 15 $M_{\odot}$ progenitor model in \citealt{Woosley:1995}.
The CCSN model employed the Togashi equation-of-state (EOS) \citep{Togashi:2017}, which is a nuclear EOS constructed by the variational many-body theory with realistic nuclear force potentials.
Following the same procedure as in \citealt{Nakazato:2018}, we remap a fluid profile (i.e.,, entropy and electron fraction profiles as functions of baryon mass) from the center to a radius just behind shock wave at 300ms after core bounce.
The corresponding baryon mass covered by the spatial grid is 1.4675$M_{\odot}$, which also represents the baryon mass of the PNS.
We construct the hydrostatic PNS structure with steady neutrino flow from this initial condition, and then simulate the following long-term evolution of the PNS.
Just as in the CCSN model, we assume that $\mu$-, $\tau$ flavors, and their antipartners are identical. This would be a reasonable condition unless on-shell muons appear \citep{Bollig:2017,Fischer:2020}.
We follow the post-explosion evolution of PNS up to 50 seconds from the remapped initial condition; the result is summarized in Fig.\,\ref{fig:mat_prof}.
In the figure, we display the radial profiles of baryon mass density ($\rho$), electron fraction ($Y_e$), temperature ($T$), and lapse function ($\alpha$) at selected time snapshots for our PNS cooling model.

MGFLD transport approximately calculates the flow of energy and lepton number, but does not provide information about the angular structure of the radiation fields needed to determine whether neutrino flavor instabilities present.
Based on the PNS model, we carry out another neutrino transport simulation for stability analysis of flavor conversion. The detail is described in the next section.

\section{Sedonu}\label{sec:Sedonu}

In general, linear stability analysis of neutrino flavor conversion requires full angular information of neutrino distributions in momentum space (see Sec.\,\ref{sec:Stab} for more details). We, hence, carry out another multi-angle neutrino transport simulation on top of the PNS fluid background.

We calculate the equilibrium neutrino radiation field on each snapshot using a general relativistic Monte Carlo neutrino transport code {\tt Sedonu} \citep{Richers:2015,Richers:2022a}.
We assume spherical symmetry and discretize the domain in radius using the same grid present in the MGFLD transport model. 
We also map the same metric, matter density, temperature, and equilibrium neutrino chemical potentials from the PNS cooling model.
At each radial grid cell, the neutrino distribution is discretized in polar angle with 100 zones uniformly distributed over $-1\leq \cos\theta_{\nu} \leq 1$.
We use the SFHo EOS \citep{Steiner:2013} and neutrino interaction rates calculated using NuLib \citep{OConnor:2015}.
We treat charged-current absorption and emission from free nucleons and nuclei and elastic isotropic scattering from nucleons and nuclei.
Emission from neutrino pair production and nucleon-nucleon bremsstrahlung radiation is included, and the inverse reactions are treated with an effective absorption opacity calculated by assuming detailed balance.
The opacities in NuLib are largely the same as those in \citealt{Bruenn:1985}, but include corrections for weak magnetism and recoil, form-factor corrections due to decoherence, electron polarization corrections, and ion-ion correlations \citep{Horowitz:1997,Horowitz:2002,Burrows:2006}.
We employ 18 energy bins with upper bounds logarithmically spaced from 2 to 311 MeV.
We treat regions with large scattering optical depths with the Monte Carlo random walk method to accelerate diffusive transport \citep{Fleck:1984}.

We simulate $2.1-4.0\times10^{10}$ Monte Carlo packets in each snapshot, each representing a number of neutrinos proportional to the total luminosity of the grid cell they are created in.
The particle positions and four-momenta move along geodesics, while the number of neutrinos contained in each packet decreases exponentially according to the absorption rate.
Monte Carlo packets change the direction after a random interval according to the scattering opacity, and are assumed to scatter isotropically in the frame comoving with the fluid.
As particles propagate, they contribute to the local distribution function.
Particles are destroyed when they leave the domain or pass within a radius of $10\,\mathrm{km}$ (although for the $30\,\mathrm{ms}$ and $50\,\mathrm{ms}$ snapshots we necessarily remove this condition to avoid artifacts associated with the finite inner boundary).
Particles are rouletted if their weight decreases below a factor of $10^{-3}$ of their original weight. 
Once all particles are destroyed, the resulting distribution is read out and then used for stability analysis of neutrino flavor conversions.

\begin{figure}
    \centering
    \includegraphics[width=\linewidth]{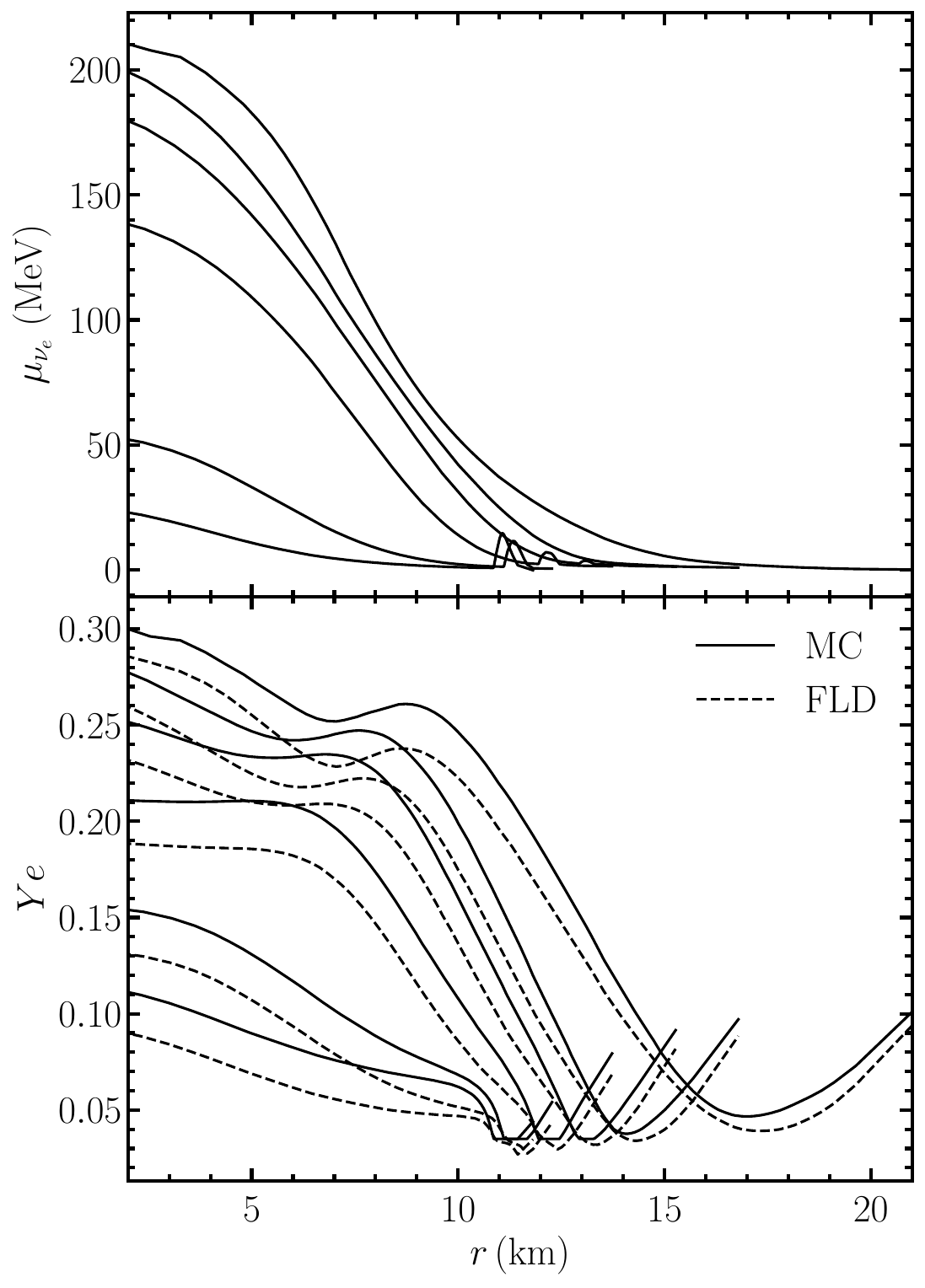}
    \caption{Equilibrium electron neutrino chemical potential and corresponding electron fraction for the FLD and MC calculations. This shows the snapshot at $1\,\mathrm{s}$.
    }
    \label{fig:matching_munue}
\end{figure}
It is important to note that neither the EOS nor the neutrino interaction rates are consistent with those used in the simulation of the PNS evolution.
Although a more exact match would be desirable, we take a more expedient first approach to understand qualitative features of the neutrino distributions.
We choose the baryon mass density, temperature, and chemical potential of $\nu_e$ ($\mu_p + \mu_e - \mu_n$) as the primitive thermodynamic variables from which to reconstruct reaction rates, and allow $Y_e$ to float to accommodate this choice.
As a result, the Monte Carlo calculation has the same equilibrium distributions as in the MGFLD calculation, even if the interaction rates and fluxes differ.
In practice, this choice also results in the similar reaction rates for weak processes. The chemical potentials of electron neutrinos and the change in $Y_e$ required are shown in Fig.\,\ref{fig:matching_munue}.

\begin{figure}
    \centering
    \includegraphics[width=\linewidth]{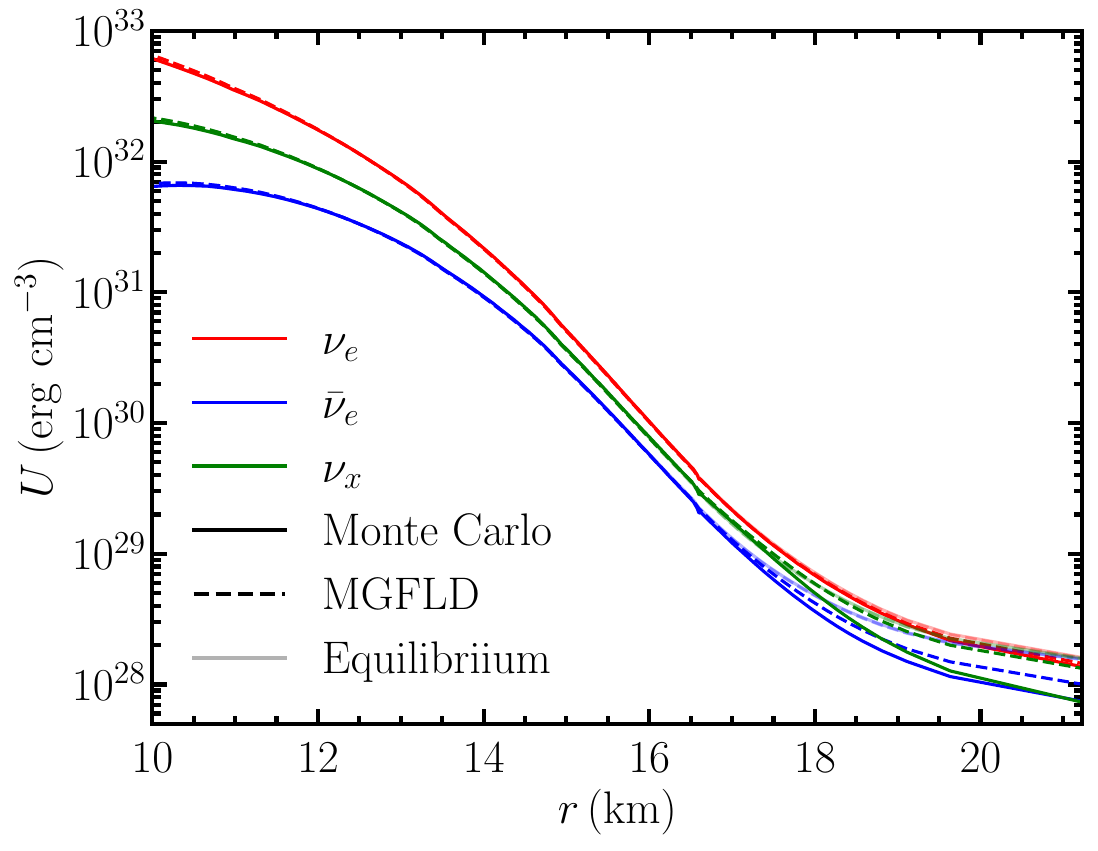}
    \caption{Net energy density in neutrinos at $1\,\mathrm{s}$.
    Solid lines are calculated by Sedonu, dashed are calculated by FLD, and transparent lines are the equilibrium distributions.}
    \label{fig:energy_density}
\end{figure}

In Fig.\,\ref{fig:energy_density}, we compare the energy density of neutrinos computed using MGFLD and {\tt Sedonu}.
Their distributions are nearly identical in optically thick region by construction, since the equilibrium chemical potentials are designed to match between the two methods.
However, they gradually deviate with increasing radius in the outer envelope of PNS.
The deviation is larger for $\bar{\nu}_e$ than $\nu_e$ due to the lower opacity of $\bar{\nu}_e$ in the more neutron rich environment of the MGFLD calculation.
One thing we do notice here is that the deviation is not only due to EOS and weak interactions but also the difference of two transport schemes.
Despite both the different transport schemes and the different opacities, the differences between the distributions of neutrino energy density are minor and the two methods are in overall good agreement with each other.
The Monte Carlo calculation, however, produces a realistic angular distribution that results from self-consistent diffusion and advection of radiation away from the PNS.

\section{Linear stability analysis}\label{sec:Stab}
By using the neutrino data obtained from {\tt Sedonu} simulations, we perform linear stability analysis with a dispersion relation approach.
Below, we provide essential details of the analysis but we refer readers to \citealt{Izaguirre:2017,Liu:2023,Akaho:2024a} for more details. Throughout this work we do not simulate neutrino quantum kinetics, but rather use the results of our classical kinetics calculations to ascertain the presence of quantum kinetic instabilities.

The fundamental quantity in neutrino quantum kinetics is neutrino density matrix ($\rho$)\footnote{The reader should not confuse $\rho$ with the baryon mass density.}, and the neutrino transport equation can be expressed in terms of $\rho$.
In our stability analysis, we work with the quantum kinetic equation (QKE) derived under the mean-field approximation, which can be described as
\begin{equation}
    (\partial_t + \boldsymbol{v}\cdot\partial_{\boldsymbol{x}})\rho = -\ii\left[\mathcal{H}, \rho\right] + \mathcal{C}, \label{eq:basicQKE}
\end{equation}
where the flavor-transforming Hamiltonian is defined in an orthonormal tetrad and is composed of vacuum, matter, and self-interaction terms:
\begin{equation}
    \mathcal{H} = U\frac{M^2}{2E}U^{\dagger} + v^{\mu}\Lambda_{\mu} + \sqrt{2}\GF \int\dd\Gamma^{\prime}\, v^{\mu}v_{\mu}^{\prime}\, \rho^{\prime}.
\end{equation}
In the first term, $U$ and $M^2$ denote the Pontecorvo-Maki-Nakagawa-Sakata matrix and the mass-squared matrix for neutrino mass eigenstates, respectively.
The second term induces matter oscillation, where $v^{\mu}=(1,\boldsymbol{v})$ is the neutrino's unit four velocity and $\Lambda_{\mu}=\sqrt{2}\GF\,\mathrm{diag}[\{j_{\mu}^{l}\}]$ is the charged lepton current $j_{\mu}^{l}$ of charged lepton $l$.
The third part corresponds to the neutrino self-interaction, and the integral over phase space is defined by
\begin{equation}
    \int\dd\Gamma \equiv \int^{\infty}_{-\infty}\frac{E_{\nu}^2\dd E_{\nu}}{2\pi^2}\int\frac{\dd\Omega_{\nu}}{4\pi}.
\end{equation}
We employ the flavor-isospin convention, i.e., $\bar{\rho}(E)\equiv -\rho(-E)$ for the anti-neutrino density matrix.
Regarding the collision term (the second term in the right hand side of Eq.\,\eqref{eq:basicQKE}), we employ the same approach as in \citealt{Liu:2023c,Akaho:2024a}, in which the emission- and absorption processes are treated in the following form;
\begin{equation}
    \mathcal{C}_{ab} = j_{a}\delta_{ab} - \left[ \langle j\rangle_{ab}+ \langle \kappa\rangle_{ab}\right]\rho_{ab}, \label{eq:ColQKE}
\end{equation}
where $j$ and $\kappa$ correspond to emissivity and absorptivity, respectively,
but we ignore scattering processes for expediency.
In the expression, the subscripts $a$ and $b$ denote a neutrino flavor, and the bracket is defined by
\begin{equation}
    \langle R\rangle_{ab} \equiv \frac{R_{a}+R_{b}}{2}.
\end{equation}
It is worth noting that pair processes are incorporated in our stability analysis under the form of Eq.\,\eqref{eq:ColQKE}.
We compute their effective emissivity $j_{\mathrm{pair}}$ and absorptivity $\kappa_{\mathrm{pair}}$ by carrying out momentum-integration of neutrino distributions with their reaction kernels (see also \citealt{Liu:2023c,Akaho:2024a}) based on the opacities used in the Monte Carlo calculations.
As a result, the collision term for heavy-leptonic neutrinos is non-zero in our stability analysis, despite no charged-current reactions.
This is in line with the approximations made in both the MGFLD and Monte Carlo calculations.

The resultant QKE for the off-diagonal component of the density matrix can be written as
\begin{equation}
\begin{split}
    \ii(\partial_t + \boldsymbol{v}\cdot\partial_{\boldsymbol{x}})\rho_{ex} &= \sqrt{2}\GF\rho_{ex}v^{\mu}\int\dd\Gamma^{\prime}\,v_{\mu}^{\prime}\,(\rho_{ee}^{\prime}-\rho_{xx}^{\prime}) \\
    &- \sqrt{2}\GF(\rho_{ee}-\rho_{xx})v^{\mu}\int\dd\Gamma^{\prime}\,v_{\mu}^{\prime}\,\rho_{ex}^{\prime} \\
    &- \ii R_{\mathrm{EA}}\rho_{ex}, \label{eq:linQKEofD}
\end{split}
\end{equation}
where emission, absorption, and pair process are summarized as
\begin{equation}
    R_{\mathrm{EA}} \equiv \left[ \langle j\rangle_{ex}+ \langle \kappa\rangle_{ex}\right].
\end{equation}
Here, the vacuum and matter terms are omitted in this equation because they do not affect FFI or CFI.
We linearize Eq.\,\eqref{eq:linQKEofD} under the condition that neutrinos are nearly in flavor eigenstates.
We then derive the dispersion relation under the assumption that the wavelength of unstable modes, if any, are sufficiently smaller than the scale height of background neutrinos.
By adopting a plane-wave ansatz, $\rho_{ex} \propto \tilde{Q}\exp[-\ii k_{\mu}x^{\mu}]$ where $k_{\mu}=(\omega,\boldsymbol{k})$, the resultant dispersion relation can be written as,
\begin{equation}
    \mathrm{det}\left[\Pi^{\mu\nu}(k)\right] = 0,
    \label{eq:DR_det}
\end{equation}
where
\begin{equation}
    \Pi^{\mu\nu} = \eta^{\mu\nu} + \sqrt{2}\GF\int\dd\Gamma\,(\rho_{ee}-\rho_{xx})\frac{v^{\mu}v^{\nu}}{v\cdot k + \ii R_{\mathrm{EA}}}.
    \label{eq:DR_Pi}
\end{equation}
If there is a positive imaginary value of $\omega$ that satisfies this condition, the neutrino distribution is unstable to flavor conversions.

Although it is straightforward to search for positive imaginary values of $\omega$ by solving Eq.\,\eqref{eq:DR_det}, we take an alternative way in the present study.
The major reason to avoid the direct search is its computational cost.
As is well known, spurious modes, corresponding to unphysical eigenmodes, appear in the stability analysis when we handle momentum-integration in Eq.\,\eqref{eq:DR_Pi} by using discretized angular distributions of neutrinos \citep{sarikas:2012,Morinaga:2018}.
We can suppress the artifact by increasing resolutions of neutrino momentum space, but neutrino distributions need to be scanned at each spatial point for multiple snapshots in the present study, exhibiting that direct method is not appropriate.
Below, we describe alternative way to identify FFI and CFI by following the procedure as in \citealt{Akaho:2024a}.

FFI is dictated by anisotropy of ELN and XLN.
Under the assumption that all heavy-leptonic neutrinos ($\mu$, $\tau$, and their antipartners) are identical, it can be determined only by ELN angular distribution, which is defined as,
\begin{equation}
    G(\Omega_{\nu}) = \int\frac{E_{\nu}^2\dd E_{\nu}}{2\pi^2}\left(f_{\nu_e}-f_{\bar{\nu}_e} \right),
\end{equation}
where $f_{\nu_a}$ is a neutrino occupation distribution for a flavor $a$, which corresponds to the diagonal component in neutrino density matrix.
As proven by \citealt{Morinaga:2022}, its zero crossings (or angular crossings) give a sufficient and necessary condition for FFI.
For this reason, we inspect ELN angular crossings in neutrino radiation fields, and then judge whether FFI can occur.

For CFI, an efficient approximate method was developed by \citealt{Liu:2023}, in which the growth rate of $\boldsymbol{k}=\boldsymbol{0}$ (or homogeneous) mode can be analytically estimated under an assumption that neutrino angular distributions are isotropic.
We note that homogeneous mode usually provides the dominant unstable mode \citep{Liu:2023}, and that the neutrino distributions inside the PNS are nearly isotropic.
It is also worth noting that the collision term of iso-energetic scattering processes become exactly zero for isotropic neutrino distributions, which also supports our neglect of scattering processes in the stability analysis.

Under these assumptions, two analytical solutions for oscillation frequency $\omega$, so-called isotropy-preserving and isotropy-breaking branches, can be given as,
\begin{equation}
    \omega_{\pm}^{\mathrm{pres}} = -A -\ii\gamma \pm \sqrt{A^2 - \alpha^2 + 2\ii G\alpha}
    \label{eq:pres_1}
\end{equation}
and
\begin{equation}
    \omega_{\pm}^{\mathrm{break}} = -\frac{A}{3} -\ii\gamma \pm \sqrt{\left(\frac{A}{3}\right)^2 - \alpha^2 - \frac{2}{3}\ii G\alpha}.
    \label{eq:break_1}
\end{equation}
Here each quantity is defined as the following notation:
\begin{equation}
    G = \frac{\mathfrak{g}+\bar{\mathfrak{g}}}{2}, A = \frac{\mathfrak{g}-\bar{\mathfrak{g}}}{2}, \gamma = \frac{\langle R\rangle+\langle \bar{R}\rangle}{2}, \alpha = \frac{\langle R\rangle-\langle \bar{R}\rangle}{2},
\end{equation}
where $\mathfrak{g} = \sqrt{2}\GF(n_{\nu_e}-n_{\nu_x})$.
The neutrino number density and mean collision rates are given by
\begin{equation}
\begin{split}
    n_{\nu_a} &= \int\dd\Gamma\, \rho_{aa} \\
    \langle R\rangle_{a} &= \frac{1}{n_{\nu_a}}\int\dd\Gamma\, R_{a}\rho_{aa}.
\end{split}
\end{equation}
We are primarily interested in determining whether Eqs.\,\eqref{eq:pres_1} and \eqref{eq:break_1} have positive imaginary parts.
Following \citealt{Liu:2023}, we can approximate Eqs.\,\eqref{eq:pres_1} and \,\eqref{eq:break_1} by taking a limit of $A$ in its relation to $\abs{G\alpha}$, which can be written as,
\begin{align}
    \mathrm{max}\left[\mathrm{Im}\,\omega^{\mathrm{pres}}\right] = 
    \begin{cases}
    -\gamma + \frac{\abs{G\alpha}}{\abs{A}} & (\mathrm{if}\,\,\, A^2 \gg \abs{G\alpha}) \\
    -\gamma + \sqrt{\abs{G\alpha}} & (\mathrm{if}\,\,\, A^2 \ll \abs{G\alpha})
    \end{cases}
    \label{eq:Imw_pres}
\end{align}
for isotropy-preserving mode and
\begin{align}
    \mathrm{max}\left[\mathrm{Im}\,\omega^{\mathrm{break}}\right] = 
    \begin{cases}
    -\gamma + \frac{\abs{G\alpha}}{\abs{A}} & (\mathrm{if}\,\,\, A^2 \gg \abs{G\alpha}) \\
    -\gamma + \frac{\sqrt{\abs{G\alpha}}}{\sqrt{3}} & (\mathrm{if}\,\,\, A^2 \ll \abs{G\alpha})
    \end{cases}
    \label{eq:Imw_break}
\end{align}
for isotropy-breaking mode.

Eqs.\,\eqref{eq:Imw_pres} and \eqref{eq:Imw_break} exhibit that the second term in the right hand side of these equations needs to overwhelm the first one ($-\gamma$) to make the system unstable.
This indicates that large neutrino-self interactions are necessary for the occurrences of CFI.
Another important remark is that, in realistic CCSN environments, the former condition ($A^2 \gg \abs{G\alpha}$) is usually satisfied.
On the other hand, the latter case ($A^2 \ll \abs{G\alpha}$) can emerge in regions with $A\sim 0$, indicating that so-called resonance-like CFI occurs \citep{Xiong:2023c,Liu:2023}.
In fact, a detailed inspection of the neutrino radiation field based on multi-dimensional CCSN simulations suggest that such a resonance-like CFI can appear in the envelope of the PNS \citep{Akaho:2024a}.
Since the growth of flavor instability could be very fast and its non-linear evolution would lead to flavor swap \citep{Kato:2024}, the CFI may play an important role on CCSN and PNS dynamics, albeit very narrow spatial region.

\section{Results}\label{sec:result}

In this study, we find no positive sign of neutrino-flavor instabilities for either FFI or CFI in any time snapshot.
Below, we discuss the physical reasons and provide some figures displaying key quantities.
It should be noted that we focus on the spatial region at $r>10\,\mathrm{km}$ hereafter.
This is because in very optically thick regions ($r<10\,\mathrm{km}$), all neutrinos are in thermal equilibrium, and both FFI and CFI can not occur in these regions\footnote{This statement would be understandable for FFI, since it is driven by anisotropic angular distributions, indicating that it can not occur in isotropic distributions as in thermal equilibrium states. For CFIs, the argument was given in our previous papers \citep{Liu:2023c,Akaho:2024a}, while we shall provide a more formal mathematical proof in another paper \citep{Liu:2024}.}.
We confirmed that $r=10\,\mathrm{km}$ corresponds to a safe radius where all neutrinos are in thermal equilibrium states at all time snapshots (see, e.g., Fig.\,\ref{fig:energy_density}).

\begin{figure}[t]
    \centering
    \includegraphics[width=0.95\linewidth]{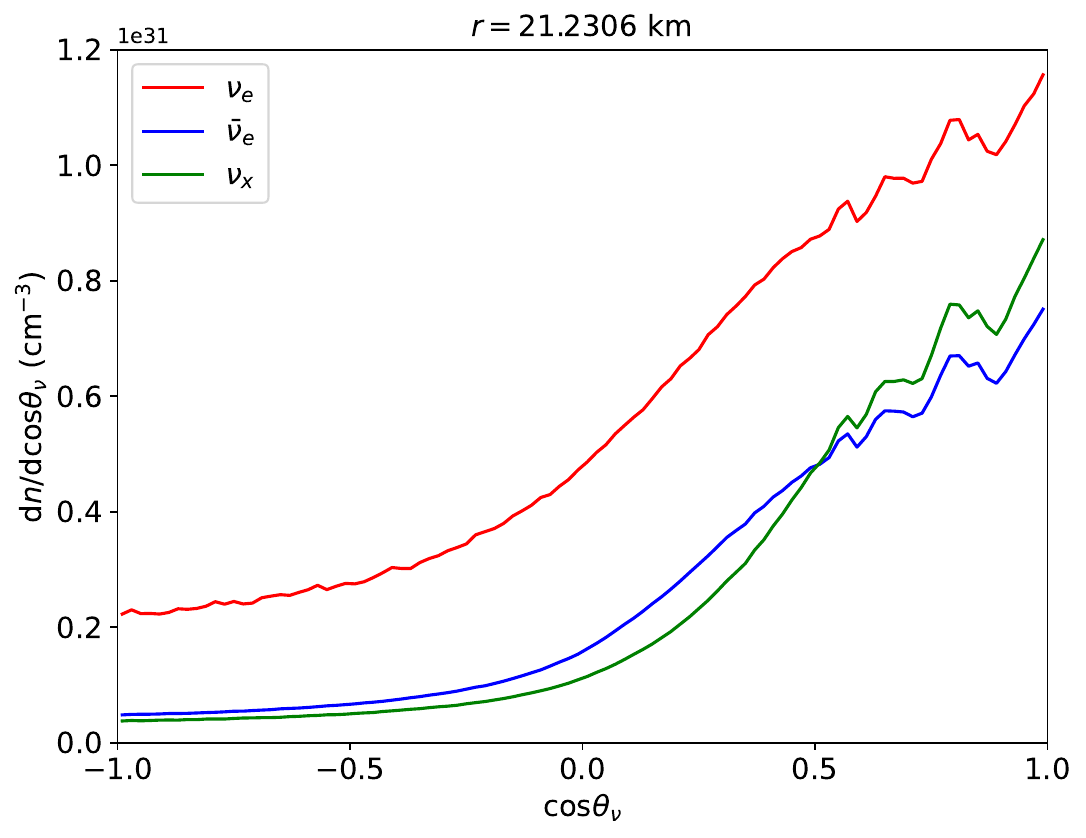}
    \caption{Angular distribution of electron neutrinos and antineutrinos at the outermost cell at $1\,\mathrm{s}$.
    There is no ELN angular crossing. A similar lack of crossing exists at every location in every snapshot.}
    \label{fig:Sedonu_ang}
\end{figure}

\begin{figure}[t]
    \centering
    \includegraphics[width=0.95\linewidth]{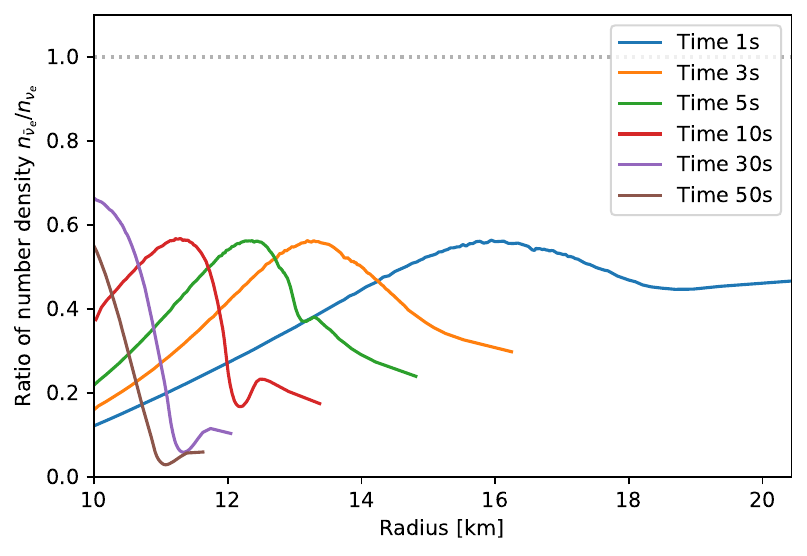}
    \caption{Radial profile of the ratio of $\bar{\nu}_e$ to $\nu_e$ in number density at each time snapshot.
    The dotted line corresponds to $n_{\bar{\nu}_e}=n_{\nu_e}$.
    }
    \label{fig:ratio_num}
\end{figure}

We begin with analyses of FFI.
Figure\,\ref{fig:Sedonu_ang} displays the angular distributions for each species as one of the representative examples in our results.
The plotted distributions are measured at the outermost cell at $1\,\mathrm{s}$ time snapshot in our PNS cooling model, and there is some noise due to the random nature of the Monte Carlo calculations.
As shown in the figure, the $\nu_e$-distribution substantially dominates over those of the other species for all angles, indicating that there are no ELN angular crossings. Although the particulars of the distribution vary with location and time, the same qualitative behavior is the same at all other locations and times.

That generation of ELN crossings is difficult in this environment can be also understood from the radial profiles for the ratio of the number density of $\bar{\nu}_e$ to $\nu_e$, shown in Fig.\,\ref{fig:ratio_num}.
In general, since the PNS is a neutron-rich environment, $\bar{\nu}_e$ is decoupled at smaller radii than $\nu_e$, resulting in more forward-peak angular distributions.
The process can lead to ELN angular crossing, but the large disparity in the number density between $\nu_e$ and $\bar{\nu}_e$ implies that a large disparity in angular distributions is required to make up for the disparity in number density to create crossings \citep{Nagakura:2019,Abbar:2019,Abbar:2020}. This never happens in our calculations.

\begin{figure}[t]
    \centering
    \includegraphics[width=0.95\linewidth]{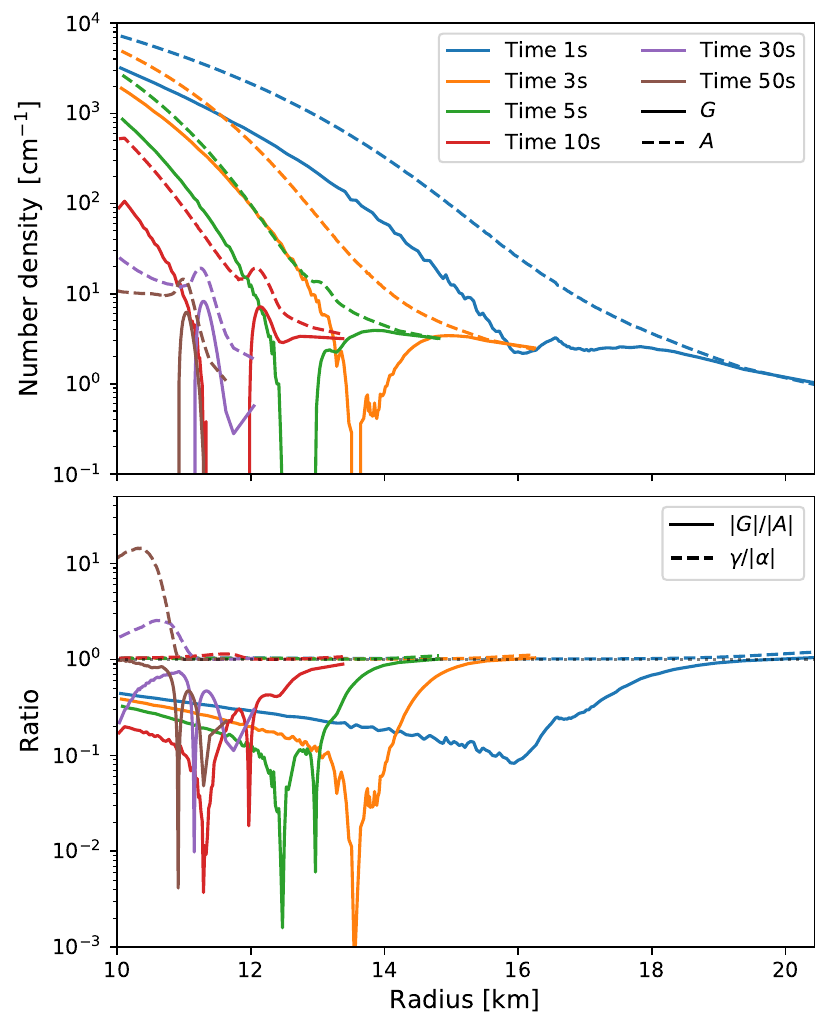}
    \caption{Radial profile of number density (top) and the ratio $\abs{G}/\abs{A}$ and $\gamma/\abs{\alpha}$ (bottom) at each time snapshot. Dotted vertical line corresponds to $\abs{G}=\abs{A}$ in the bottom panel.}
    \label{fig:growth_rate}
\end{figure}

Let us turn our attention to the CFI.
The rationale behind the stable neutrino radiation field with respect to CFI is interpretable with analytical formulae (see Eqs.\,\eqref{eq:Imw_pres} and \eqref{eq:Imw_break}).
The condition for unstable modes is that the second term $\abs{G\alpha}/\abs{A}$ overwhelms the first one, $\gamma$, in Eqs.\,\eqref{eq:Imw_pres} and \,\eqref{eq:Imw_break} for non-resonant cases ($A^2 \gg \abs{G\alpha}$)\footnote{We also note that there are no regions with $A^2 \ll \abs{G\alpha}$ in this study. This exhibits that it does not meet a condition for resonance-like CFI.}.
The condition can be simplified into $\abs{G}/\abs{A} > \gamma/\abs{\alpha}$.
Since $\gamma$ is larger than $\abs{\alpha}$, we further simplify the inequality as $\abs{G}/\abs{A} > 1$, which corresponds to a necessary (though not sufficient) condition for the occurrences of CFIs.
In Fig.\,\ref{fig:growth_rate}, we portray the radial profiles of $G$ and $A$ in the top panel, while those for $\abs{G}/\abs{A}$ and $\gamma/\abs{\alpha}$ are displayed in the bottom.
As shown in the bottom panel, $\abs{G}/\abs{A}$ is always lower than unity, exhibiting that no CFI occurs in the entire spatial region at all time snapshots.

\begin{figure}[t]
    \centering
    \includegraphics[width=0.95\linewidth]{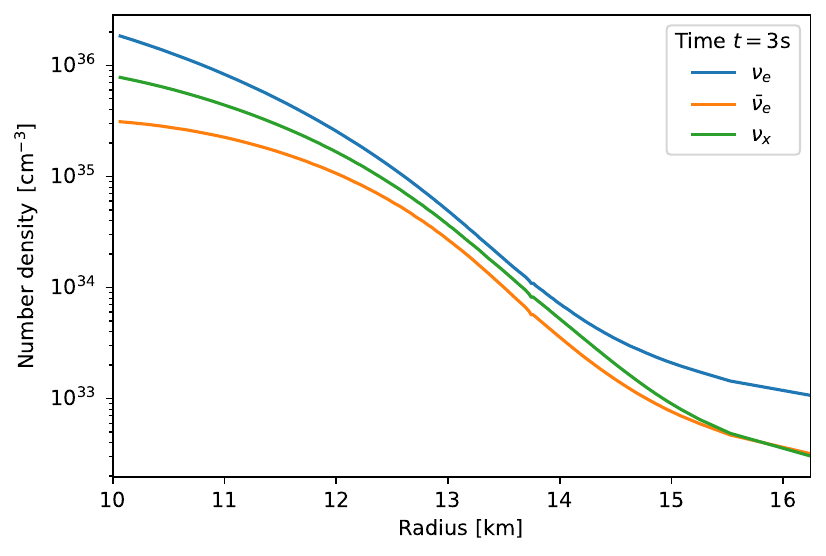}
    \includegraphics[width=0.95\linewidth]{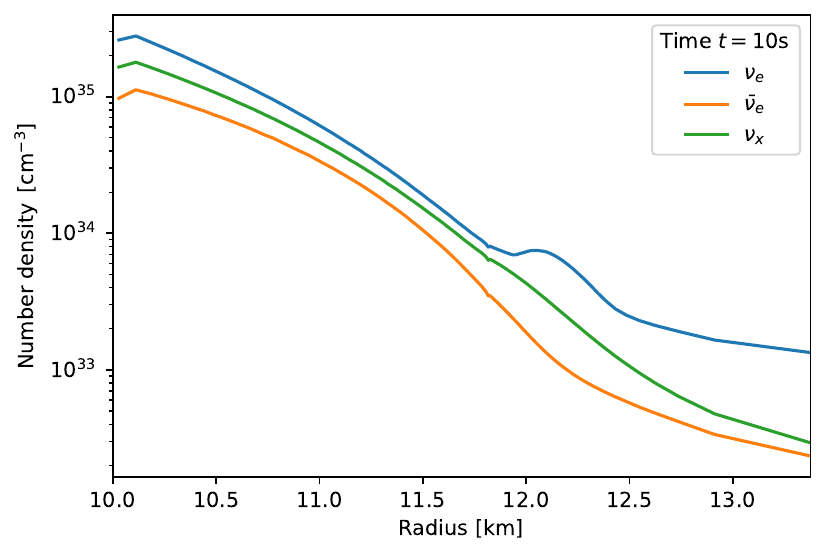}
    \caption{Neutrino number density for $\nu_e$ (blue), $\bar{\nu}_e$ (orange), and $\nu_x$ (green, one species) at time snapshots $t=3$ and $10\,\mathrm{s}$.}
    \label{fig:nu_num}
\end{figure}

\begin{figure}[t]
    \centering
    \includegraphics[width=0.95\linewidth]{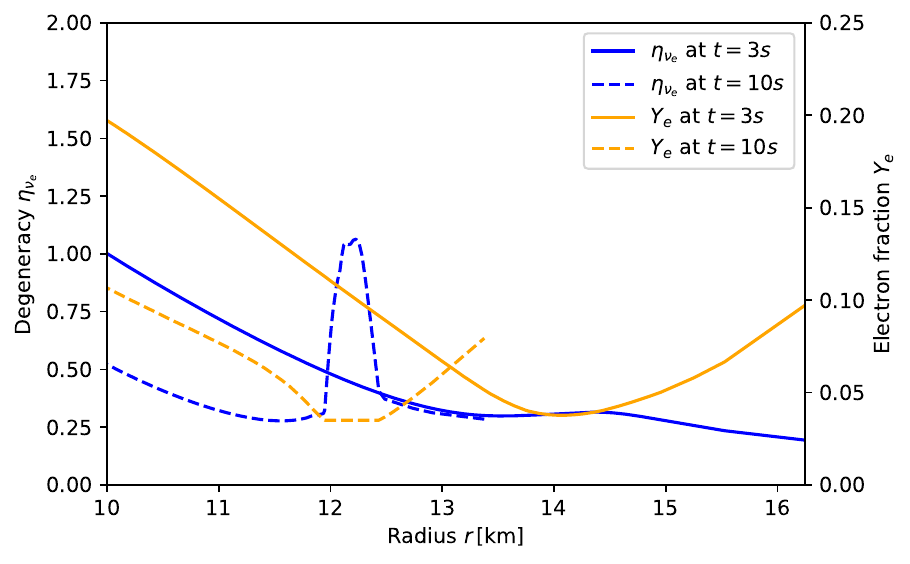}
    \caption{Radial profile for degeneracy of electron neutrinos $\eta_{\nu_e}$ (blue) and electron fraction $Y_e$ (orange).
    }
    \label{fig:Ye_deg}
\end{figure}

It is worth noting that the small $G$ is attributed to the large populations of $n_{\nu_x}$.
In fact, the inequality $\abs{G}/\abs{A} > 1$ is satisfied when both $\nu_e$ and $\bar{\nu}_e$ are larger than $\nu_x$\footnote{We note that the inequality is also satisfied in cases where $\nu_x$ is much larger than $\nu_e$ and $\bar{\nu}_e$. However, such a situation does not arise in CCSN environments.}.
It means that $\nu_x$ suppresses the appearance of CFI, as reported by our previous paper \citep{Liu:2023c}.
To see the trend clearly, we provide Fig.\,\ref{fig:nu_num}, in which the radial profiles of neutrino number density for each species are portrayed at two time snapshots $t=3$ and $10\,\mathrm{s}$, as examples.
As shown in the figure, the $\nu_x$ number densities always dominate over those of $\bar{\nu}_e$, which results in $\abs{G}/\abs{A}<1$.
To understand the trend of relatively low densities of $\bar{\nu}_e$ compared to other species of neutrinos, we depict the radial profiles of degeneracy of electron neutrinos, which is defined as $\eta_{\nu_e}\equiv \mu_{\nu_e}/T$, and also $Y_e$ in Fig.\,\ref{fig:Ye_deg}.
The positive value of $\eta$ suggests that Fermi degeneracy of $\nu_e$ accounts for the suppression of $\bar{\nu}_e$.

We remark that the peak profile of $\eta_{\nu_e}$ around $r=12\,\mathrm{km}$ at $t=10\,\mathrm{s}$ might be a numerical artifact in the PNS cooling model.
In fact, the diffusion approximation for neutrino transport is not reasonable in the region.
It should be also mentioned that the matter distributions around the surface of PNS is sensitive to fall-back or long-lasting accretion flows onto the PNS \citep{Nagakura:2021d,Akaho:2024}.
We should keep in mind these uncertainties, and the result might be changed in more self-consistent PNS cooling models.

\begin{figure}[t]
    \centering
    \includegraphics[width=0.95\linewidth]{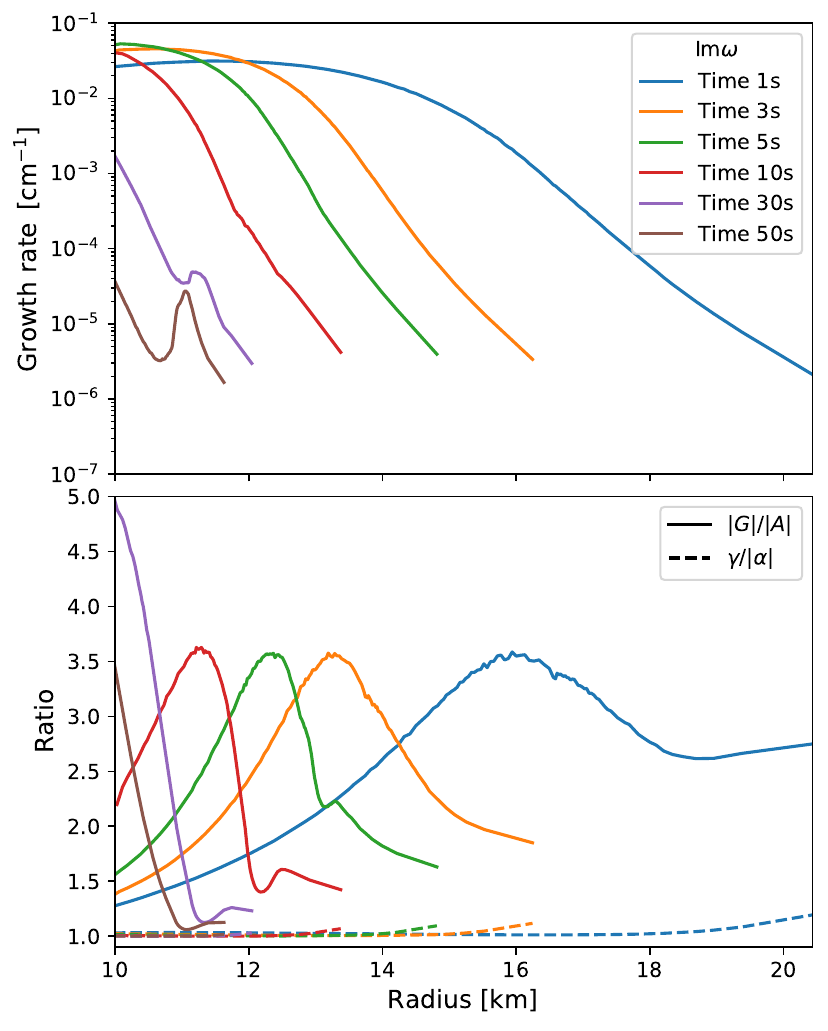}
    \caption{Same as Fig.\,\ref{fig:growth_rate}, but for without contributions from heavy-leptonic flavors.
    }
    \label{fig:no_nuX}
\end{figure}

We provide another piece of evidence that $\nu_x$ is responsible for suppressing CFI.
Figure\,\ref{fig:no_nuX} shows the growth rates and the ratios without the contributions from $\nu_x$ (more specifically, we set $n_{\nu_x}=n_{\bar{\nu}_x}=0$ in computing $G$ and $A$).
As shown clearly in the top panel, positive growth rates appear.
Also, in the bottom panel of the figure, $G$ becomes substantially larger than that in the cases with $\nu_x$ (see also Fig.\,\ref{fig:growth_rate}).
On the other hand, $A$ remains the same as the case with $\nu_x$.
This is simply because $A$ does not depend on $\nu_x$, as long as $\nu_x = \bar{\nu}_x$ is satisfied.

Our result shows the importance of accurate modeling of $\nu_x$ radiation fields to assess the occurrence of CFIs.
It should be mentioned that there are large uncertainties in weak processes of $\nu_x$ even for modern CCSN models, which are, for instance, many-body corrections in both neutral- and charged-current reactions \citep{Burrows:1998,Burrows:1999,Horowitz:2017}, nucleon-nucleon bremsstrahlung \citep{Friman:1979,Bartl:2016,Guo:2019}, and on-shell muons \citep{Fischer:2020,Guo:2020,Sugiura:2022}, etc. 
We, thus, encourage careful consideration of how these rates can have an influence on CFIs, whose detailed investigations are deferred to future work.

\section{Conclusions and summary}\label{sec:conclusion}
In this paper, we investigate the possibilities of flavor instabilities during the PNS cooling phase.
To this end, we construct a spherically symmetric PNS model by employing a quasi-static approximation and a flux-limited diffusion treatment for neutrino transport.
For an accurate assessment of neutrino-flavor instabilities, we run another multi-angle neutrino transport simulation by {\tt Sedonu} on top of fluid distributions obtained from the PNS cooling model.
We note that full angular information on neutrinos is necessary to assess flavor instabilities, particularly for the FFI.
According to our analyses, we find no positive sign of either FFI or CFI in our model.

We also discuss the physical reasons for the absence of FFI and CFI.
For both FFI and CFI, high degeneracy of $\nu_e$ weakens the emission of $\bar{\nu}_e$ and hampers the appearance of flavor instabilities.
We also find that CFI is sensitive to $\nu_x$.
As an evidence, we show that CFI can occur if there is no contribution from $\nu_x$.
We, thus, conclude that the combined effects of $\nu_x$ and the weak emission of $\bar{\nu}_e$ account for suppressing CFI.

Before we close this paper, we describe some important limitations in the present study.
Most importantly, we neglect multi-dimensional effects such as PNS convection.
As shown in previous studies, the results of flavor instabilities can be very sensitive to the dimensionality; in fact, FFIs tend to be suppressed in spherically symmetric models, but multi-dimensional fluid instabilities offer environments more conducive to development of the FFI.
This is mainly because PNS convection can facilitate deleptonization of CCSN core \citep{Dessart:2006,Buras:2006,Nagakura:2020}, that reduces the disparity between $\nu_e$ and $\bar{\nu}_e$.
As a result, ELN angular crossings can occur much more easily than in spherically symmetric models \citep{DelfanAzari:2020,Glas:2020,Nagakura:2021b,Akaho:2023}.
We also note that reducing the lepton number in the core would facilitate resonance-like CFI (see \citealt{Akaho:2024a}).
On the other hand, PNS convection can increase $\nu_x$ diffusion \citep{Nagakura:2020}, which may work to suppress CFIs.
This suggests that PNS convection has two competing effects regarding CFIs, exhibiting requirements of detailed investigations.

Another issue in the present study is neutrino-matter interactions.
As described in Sec.\,\ref{sec:result}, neutrino-matter interactions are one of the most important ingredients to determine neutrino radiation fields, but our current treatments contain an incomplete and imperfect set of neutrino interaction processes, and the dynamical diffusion transport uses interaction rates that are different from the post-processed Monte Carlo calculation.
In the future, it will be important to do similar analyses with high-fidelity input physics in order to draw more robust conclusions about whether flavor instabilities can occur and have the ability to impact neutrino signals, nucleosynthesis, and PNS cooling.
The results in this study serve as an important reference for future work.

\section{Acknowledgments}
M.Z. is supported by the Japan Society for Promotion of Science (JSPS) Grant-in-Aid for JSPS Fellows (Grants No. 22KJ2906) from the Ministry of Education, Culture, Sports, Science, and Technology (MEXT) in Japan.
H.N. is supported by Grant-in-Aid for Scientific Research (23K03468) and also by the NINS International Research Exchange Support Program.
S.R. was supported by a National Science Foundation Astronomy and Astrophysics Postdoctoral Fellowship under award AST-2001760.
H.S. is supported by JSPS Grant-in-Aid for Scientific Research on Innovative Areas ``Unraveling the History of the Universe and Matter Evolution with Underground Physics'' (No. 19H05802 and No. 19H05811) and for Transformative Research Areas ``The creation of multi-messenger astrophysics'' (No. 24H01817).
C.K. is supported by JSPS KAKENHI Grant Numbers JP20K14457, JP22H04577 and JP24H02245.
The numerical computations were partly carried out on Cray XC50 at the Center for Computational Astrophysics in National Astronomical Observatory of Japan.
This work is also supported by the HPCI System Research Project (Project ID:240079).



\appendix

\section{Numerical scheme for quasi-static PNS evolution}\label{Ap:PNScoolingscheme}
Our numerical scheme for the quasistatic PNS cooling can be summarized as follows.
We use the metric for the spherical space-time:
\begin{eqnarray}
  \dd s^{2}
  &=& \ee^{2\phi}(c\dd t)^{2} - \frac{1}{\Gamma^2}\dd r^2
       - r^{2}(\dd\theta^{2} + \sin^{2} \theta \dd\varphi^2 )\nonumber\\
  &=& \ee^{2\phi}(c\dd t)^{2} - \left(\frac{\dd m_B}{4\pi r^2\rho_B} \right)^{2}
       - r^{2}(\dd\theta^{2} + \sin^{2} \theta \dd\varphi^2 )~~,
\end{eqnarray}
where $t$ is the time at infinity, $m_B$ is the total baryon mass inside the sphere whose area is $4 \pi r^{2}$:
\begin{eqnarray}
&&
  m_B = \int^{r}_{0} \frac{1}{\Gamma} 4 \pi r^{2} \rho_B \dd r ~~,
\end{eqnarray}
and the gravitational mass $m_G$ is
\begin{eqnarray}
&&
  m_G = \int^{r}_{0} 4 \pi r^{2} \left(\rho_B+\frac{u}{c^2}\right) \dd r ~~,
\end{eqnarray}
where $\rho_B=m_u n_B$ is the baryon (mass) density, and $m_{u}$, $n_B$, $u=u_{m}+u_{\nu}$ are atomic mass unit, baryon number density, total internal energy density of the matter and neutrinos, respectively.

The hydrostatic structure of the PNS can be derived from equations:
\begin{eqnarray}
 \frac{\partial \phi}{\partial m_B} &=& - \frac{1}{ \rho_B c^{2} + u +
  p } \frac{\partial p}{\partial m_B} \\
 \frac{\partial p}{\partial m_B} &=& - \frac{G}{r^{2}} 
   \frac{ ( \rho_B + \frac{u+p}{c^{2}} )( m_G + \frac{4 \pi r^{3}
       p}{c^{2}} )}{4 \pi r^{2} \rho_B \Gamma} ~~\mbox{(TOV equation)}
 \\ 
 \Gamma &=& \sqrt{ 1 - \frac{2 G m_G}{rc^{2}} }~~,
\end{eqnarray}
where $p \equiv p_{m} + p_{\nu}$ is the total pressure of the matter and the neutrinos.

As for the differential equation of $\phi$, the following boundary condition is used so that the metric can be connected smoothly to the Schwarzschild metric outside the star:
\begin{eqnarray}
    \ee^{\phi} = \sqrt{ 1 - \frac{2GM_G}{Rc^{2}} } ~~{\rm at~surface}~~,
\end{eqnarray}
where $M_G$ is the total gravitational mass of the PNS and $R$ is the surface radius.
$M_G$ and $R$ are determined by solving the TOV equation with the given total baryon mass of the PNS.

As for neutrino transfer, we use multi-group flux limited diffusion scheme.
The neutrino number density per unit energy $n_{\nu}(\omega)$ is related to the zero-th angular moment of the neutrino distribution function ($f_{\nu}$) as
\begin{eqnarray}
 n_{\nu}(\omega) \equiv \frac{1}{(2\pi\hbar c)^3} \int f_{\nu} \omega^{2} \dd\Omega 
\end{eqnarray}
where $\omega$ is the neutrino energy and $\int \dd\Omega$ means the angular integration.
The time evolution of $n_{\nu}(\omega)$ is caused by advection, compression and weak interactions with the matter.
The neutrino number change due to the advection can be expressed as the gradient of the neutrino number flux per unit energy $F_{\nu}(\omega)$.
In addition we introduce the red-shifted energy, $\omega_0$, which is given by
\begin{eqnarray}
 \omega_0 \equiv \omega \ee^{\phi(t,m_B)} ~~.
\end{eqnarray}
The neutrino which was emitted at $m_B$ with the energy $\omega$ is red-shifted to the energy $\omega_0$ for an observer at infinity.
We use $t$, $m_B$ and $\omega_0$ as the independent variables.

The transfer equation becomes
\begin{eqnarray}
&& \left( \frac{\partial}{\partial t_p} \frac{n_{\nu}(\omega)}{\rho_B} \right)_{m_B,\omega_0} 
 +  \frac{1}{3} \left( \frac{\partial\ln \rho_B}{\partial t_p} \right)_{m_B} \cdot 
   \left( \frac{\partial }{\partial\omega_0} 
        \left( \omega_0 \frac{n_{\nu}(\omega)}{\rho_B} \right) \right)_{t,m_B} \nonumber \\ 
 &+& \left( \frac{\partial}{\partial m_B} 4 \pi r^{2} F_{\nu}(\omega) \right)_{t,\omega_0}
 = \frac{1}{\rho_B} \frac{c}{(2\pi\hbar c)^3} \int {\rm Coll}[f_{\nu}] \omega^{2} \dd\Omega~,
\end{eqnarray}
where $\dd t_p=\ee^{\phi}\dd t$ denotes the proper time interval at $m_B$ corresponding to the time interval at infinity, $\dd t$.
Note that this equation contains the general relativistic effects such as the time dilation and the red shift.
This equation tells us how the neutrino distribution per unit mass changes.
The second term of the left hand side expresses the spectral change caused by the matter expansion and compression.
The third term denotes the neutrino flow relative to the matter, clearly.
The right hand side (the collision term) corresponds to the neutrino absorption, emission and scattering by means of the weak interaction with the matter.
The collision term of the Boltzmann equation can be expressed schematically in the form of 
\begin{eqnarray}
 {\rm Coll}[f_{\nu}] = - \frac{1}{\lambda_{\nu}} f_{\nu} + j_{\nu} ~~,
\end{eqnarray}
where $\lambda_{\nu}$ is the neutrino mean free path and $j_{\nu}$ is the source function.

In the diffusion approximation, the neutrino flux is proportional to the gradient of the neutrino density as
\begin{eqnarray}
 F_{\nu}(\omega)
            & = & - \frac{c}{3} \lambda_{\nu} \ee^{-2\phi} \Gamma 
  \left( \frac{\partial}{\partial r} n_{\nu}(\omega) \ee^{2\phi} \right)_{t,\omega_0}~~.
\end{eqnarray}
Since this flux diverges in the transparent region ($\lambda_{\nu}\rightarrow\infty$), we introduce a flux limiter $\Lambda_{\nu}$  which literally limits the outgoing flux in place of $\lambda_{\nu}$.
Among various flux limiters proposed, we adopt one used by \citealt{Mayle:1987} in this study.
\begin{eqnarray}
 \Lambda_{\nu} \equiv \frac{3}{3+x \left(1+\frac{3}{1+\frac{x}{2}+\frac{x^{2}}{8}} \right) }
 \cdot \lambda_{\nu}~~,
\end{eqnarray}
where
\begin{eqnarray}
 x \equiv \lambda_{\nu} \frac{ \left| \Gamma \frac{\partial
     n_{\nu}(\omega)\ee^{2\phi}}{\partial r} \right| }{n_{\nu}(\omega)\ee^{2\phi}} ~~.
\end{eqnarray}
The neutrino flux is re-defined as
\begin{eqnarray}
 F_{\nu}(\omega)
            & = & - \frac{c}{3} \Lambda_{\nu} \ee^{-2\phi} \Gamma 
  \left( \frac{\partial}{\partial r} n_{\nu}(\omega) e^{2\phi}
  \right)_{t,\omega_0} ~~.
\end{eqnarray}

When neutrinos interact with the matter, the electron fraction $Y_e$ and the matter entropy change due to the exchange of the electron-type lepton numbers and energies between the neutrinos and the matter.
The equation for $Y_e$ becomes
\begin{eqnarray}
 \frac{\partial Y_e}{\partial t_p}
  & = & - \frac{1}{n_B} \frac{c}{(2\pi\hbar c)^3} 
        \left( \int {\rm Coll} [f_{\nu_e}] \omega^{2} \dd\omega \dd\Omega 
             - \int {\rm Coll} [f_{\bar{\nu}_e}] \omega^{2} \dd\omega \dd\Omega \right) ~~.
\end{eqnarray}
and the equation of the entropy change becomes
\begin{eqnarray}
 T \frac{\partial s}{\partial t_p} = 
 & - & \frac{1}{n_B} \sum_{\nu} \frac{c}{(2\pi \hbar c)^3} 
   \int (\omega -\mu_{\nu}) {\rm Coll}[f_{\nu}] \omega^{2} \dd\omega \dd\Omega ~~,
\end{eqnarray}
where $T$ is the temperature and $\mu_{\nu}$ is the neutrino chemical potential for neutrinos in $\beta$ equilibrium with the matter (protons, neutrons, electrons, photons):
\begin{eqnarray}
 \mu_{\bar{\nu}_e} = - \mu_{\nu_e} =-( \mu_p+\mu_{e^-}-\mu_n) ~,~~ 
 \mu_{\nu_{\mu}}= \mu_{\bar{\nu}_{\mu}}=
 \mu_{\nu_{\tau}}= \mu_{\bar{\nu}_{\tau}}= 0 ~~.
\end{eqnarray}
With the above equations, we calculate the time evolution of $\rho_B(m_B)$, $r(m_B)$, $s(m_B)$, $T(m_B)$, $Y_e(m_B)$, $n_{\nu}(\omega,m_B)$, $\phi(m_B)$ simultaneously using implicit scheme for a very long time ($O(100)$s).
The numerical error for the total energy ($M_Gc^2+\sum_{\nu}\int L_{\nu}\dd t$) is less than 0.01\%.

\bibliography{paper}{}

\begin{thebibliography}{}
\expandafter\ifx\csname natexlab\endcsname\relax\def\natexlab#1{#1}\fi
\providecommand{\url}[1]{\href{#1}{#1}}
\providecommand{\dodoi}[1]{doi:~\href{http://doi.org/#1}{\nolinkurl{#1}}}
\providecommand{\doeprint}[1]{\href{http://ascl.net/#1}{\nolinkurl{http://ascl.net/#1}}}
\providecommand{\doarXiv}[1]{\href{https://arxiv.org/abs/#1}{\nolinkurl{https://arxiv.org/abs/#1}}}

\bibitem[{Abbar {et~al.}(2019)Abbar, Duan, Sumiyoshi, Takiwaki, \& Volpe}]{Abbar:2019}
Abbar, S., Duan, H., Sumiyoshi, K., Takiwaki, T., \& Volpe, M.~C. 2019, Physical Review D, 100, 043004, \dodoi{10.1103/PhysRevD.100.043004}

\bibitem[{Abbar {et~al.}(2020)Abbar, Duan, Sumiyoshi, Takiwaki, \& Volpe}]{Abbar:2020}
---. 2020, Physical Review D, 101, 043016, \dodoi{10.1103/PhysRevD.101.043016}

\bibitem[{Akaho {et~al.}(2024{\natexlab{a}})Akaho, Liu, Nagakura, Zaizen, \& Yamada}]{Akaho:2024a}
Akaho, R., Liu, J., Nagakura, H., Zaizen, M., \& Yamada, S. 2024{\natexlab{a}}, Physical Review D, 109, 023012, \dodoi{10.1103/PhysRevD.109.023012}

\bibitem[{Akaho {et~al.}(2024{\natexlab{b}})Akaho, Nagakura, \& Foglizzo}]{Akaho:2024}
Akaho, R., Nagakura, H., \& Foglizzo, T. 2024{\natexlab{b}}, The Astrophysical Journal, 960, 116, \dodoi{10.3847/1538-4357/ad118c}

\bibitem[{Akaho {et~al.}(2023)Akaho, Harada, Nagakura, Iwakami, Okawa, Furusawa, Matsufuru, Sumiyoshi, \& Yamada}]{Akaho:2023}
Akaho, R., Harada, A., Nagakura, H., {et~al.} 2023, The Astrophysical Journal, 944, 60, \dodoi{10.3847/1538-4357/acad76}

\bibitem[{Bartl {et~al.}(2016)Bartl, Bollig, Janka, \& Schwenk}]{Bartl:2016}
Bartl, A., Bollig, R., Janka, H.-T., \& Schwenk, A. 2016, Physical Review D, 94, 083009, \dodoi{10.1103/PhysRevD.94.083009}

\bibitem[{Bollig {et~al.}(2017)Bollig, Janka, Lohs, {Mart{\'i}nez-Pinedo}, Horowitz, \& Melson}]{Bollig:2017}
Bollig, R., Janka, H.-T., Lohs, A., {et~al.} 2017, Physical Review Letters, 119, 242702, \dodoi{10.1103/PhysRevLett.119.242702}

\bibitem[{Bruenn(1985)}]{Bruenn:1985}
Bruenn, S.~W. 1985, The Astrophysical Journal Supplement Series, 58, 771, \dodoi{10.1086/191056}

\bibitem[{Buras {et~al.}(2006)Buras, Janka, Rampp, \& Kifonidis}]{Buras:2006}
Buras, R., Janka, H.-T., Rampp, M., \& Kifonidis, K. 2006, Astronomy \& Astrophysics, 457, 281, \dodoi{10.1051/0004-6361:20054654}

\bibitem[{Burrows \& Lattimer(1986)}]{Burrows:1986}
Burrows, A., \& Lattimer, J.~M. 1986, The Astrophysical Journal, 307, 178, \dodoi{10.1086/164405}

\bibitem[{Burrows {et~al.}(2020)Burrows, Radice, Vartanyan, Nagakura, Skinner, \& Dolence}]{Burrows:2020}
Burrows, A., Radice, D., Vartanyan, D., {et~al.} 2020, Monthly Notices of the Royal Astronomical Society, 491, 2715, \dodoi{10.1093/mnras/stz3223}

\bibitem[{Burrows {et~al.}(2006)Burrows, Reddy, \& Thompson}]{Burrows:2006}
Burrows, A., Reddy, S., \& Thompson, T.~A. 2006, Nuclear Physics A, 777, 356, \dodoi{10.1016/j.nuclphysa.2004.06.012}

\bibitem[{Burrows \& Sawyer(1998)}]{Burrows:1998}
Burrows, A., \& Sawyer, R.~F. 1998, Physical Review C, 58, 554, \dodoi{10.1103/PhysRevC.58.554}

\bibitem[{Burrows \& Sawyer(1999)}]{Burrows:1999}
---. 1999, Physical Review C, 59, 510, \dodoi{10.1103/PhysRevC.59.510}

\bibitem[{Burrows \& Vartanyan(2021)}]{Burrows:2021}
Burrows, A., \& Vartanyan, D. 2021, Nature, 589, 29, \dodoi{10.1038/s41586-020-03059-w}

\bibitem[{Capozzi \& Saviano(2022)}]{Capozzi:2022}
Capozzi, F., \& Saviano, N. 2022, Universe, 8, 94, \dodoi{10.3390/universe8020094}

\bibitem[{Dasgupta {et~al.}(2017)Dasgupta, Mirizzi, \& Sen}]{Dasgupta:2017}
Dasgupta, B., Mirizzi, A., \& Sen, M. 2017, Journal of Cosmology and Astroparticle Physics, 2017, 019, \dodoi{10.1088/1475-7516/2017/02/019}

\bibitem[{Delfan~Azari {et~al.}(2020)Delfan~Azari, Yamada, Morinaga, Nagakura, Furusawa, Harada, Okawa, Iwakami, \& Sumiyoshi}]{DelfanAzari:2020}
Delfan~Azari, M., Yamada, S., Morinaga, T., {et~al.} 2020, Physical Review D, 101, 023018, \dodoi{10.1103/PhysRevD.101.023018}

\bibitem[{Dessart {et~al.}(2006)Dessart, Burrows, Livne, \& Ott}]{Dessart:2006}
Dessart, L., Burrows, A., Livne, E., \& Ott, C.~D. 2006, The Astrophysical Journal, 645, 534, \dodoi{10.1086/504068}

\bibitem[{Duan {et~al.}(2006)Duan, Fuller, Carlson, \& Qian}]{Duan:2006b}
Duan, H., Fuller, G.~M., Carlson, J., \& Qian, Y.-Z. 2006, Physical Review D, 74, 105014, \dodoi{10.1103/PhysRevD.74.105014}

\bibitem[{Ehring {et~al.}(2023{\natexlab{a}})Ehring, Abbar, Janka, Raffelt, \& Tamborra}]{Ehring:2023}
Ehring, J., Abbar, S., Janka, H.-T., Raffelt, G., \& Tamborra, I. 2023{\natexlab{a}}, Physical Review D, 107, 103034, \dodoi{10.1103/PhysRevD.107.103034}

\bibitem[{Ehring {et~al.}(2023{\natexlab{b}})Ehring, Abbar, Janka, Raffelt, \& Tamborra}]{Ehring:2023a}
---. 2023{\natexlab{b}}, Physical Review Letters, 131, 061401, \dodoi{10.1103/PhysRevLett.131.061401}

\bibitem[{Fischer {et~al.}(2020{\natexlab{a}})Fischer, Guo, Dzhioev, {Mart{\'i}nez-Pinedo}, Wu, Lohs, \& Qian}]{Fischer:2020a}
Fischer, T., Guo, G., Dzhioev, A.~A., {et~al.} 2020{\natexlab{a}}, Physical Review C, 101, 025804, \dodoi{10.1103/PhysRevC.101.025804}

\bibitem[{Fischer {et~al.}(2024)Fischer, Guo, Langanke, {Mart{\'i}nez-Pinedo}, Qian, \& Wu}]{Fischer:2024}
Fischer, T., Guo, G., Langanke, K., {et~al.} 2024, Progress in Particle and Nuclear Physics, 137, 104107, \dodoi{10.1016/j.ppnp.2024.104107}

\bibitem[{Fischer {et~al.}(2020{\natexlab{b}})Fischer, Guo, {Mart{\'i}nez-Pinedo}, Liebend{\"o}rfer, \& Mezzacappa}]{Fischer:2020}
Fischer, T., Guo, G., {Mart{\'i}nez-Pinedo}, G., Liebend{\"o}rfer, M., \& Mezzacappa, A. 2020{\natexlab{b}}, Physical Review D, 102, 123001, \dodoi{10.1103/PhysRevD.102.123001}

\bibitem[{Fleck \& Canfield(1984)}]{Fleck:1984}
Fleck, J.~A., \& Canfield, E.~H. 1984, Journal of Computational Physics, 54, 508, \dodoi{10.1016/0021-9991(84)90130-X}

\bibitem[{Friman \& Maxwell(1979)}]{Friman:1979}
Friman, B.~L., \& Maxwell, O.~V. 1979, The Astrophysical Journal, 232, 541, \dodoi{10.1086/157313}

\bibitem[{Glas {et~al.}(2020)Glas, Janka, Capozzi, Sen, Dasgupta, Mirizzi, \& Sigl}]{Glas:2020}
Glas, R., Janka, H.-T., Capozzi, F., {et~al.} 2020, Physical Review D, 101, 063001, \dodoi{10.1103/PhysRevD.101.063001}

\bibitem[{Guo \& {Mart{\'i}nez-Pinedo}(2019)}]{Guo:2019}
Guo, G., \& {Mart{\'i}nez-Pinedo}, G. 2019, The Astrophysical Journal, 887, 58, \dodoi{10.3847/1538-4357/ab536d}

\bibitem[{Guo {et~al.}(2020)Guo, {Mart{\'i}nez-Pinedo}, Lohs, \& Fischer}]{Guo:2020}
Guo, G., {Mart{\'i}nez-Pinedo}, G., Lohs, A., \& Fischer, T. 2020, Physical Review D, 102, 023037, \dodoi{10.1103/PhysRevD.102.023037}

\bibitem[{Horowitz(1997)}]{Horowitz:1997}
Horowitz, C.~J. 1997, Physical Review D, 55, 4577, \dodoi{10.1103/PhysRevD.55.4577}

\bibitem[{Horowitz(2002)}]{Horowitz:2002}
---. 2002, Physical Review D, 65, 043001, \dodoi{10.1103/PhysRevD.65.043001}

\bibitem[{Horowitz {et~al.}(2017)Horowitz, Caballero, Lin, O'Connor, \& Schwenk}]{Horowitz:2017}
Horowitz, C.~J., Caballero, O.~L., Lin, Z., O'Connor, E., \& Schwenk, A. 2017, Physical Review C, 95, 025801, \dodoi{10.1103/PhysRevC.95.025801}

\bibitem[{Izaguirre {et~al.}(2017)Izaguirre, Raffelt, \& Tamborra}]{Izaguirre:2017}
Izaguirre, I., Raffelt, G., \& Tamborra, I. 2017, Physical Review Letters, 118, 021101, \dodoi{10.1103/PhysRevLett.118.021101}

\bibitem[{Janka(2012)}]{Janka:2012a}
Janka, H.-T. 2012, Annual Review of Nuclear and Particle Science, 62, 407, \dodoi{10.1146/annurev-nucl-102711-094901}

\bibitem[{Janka(2017)}]{Janka:2017}
---. 2017, in Handbook of {{Supernovae}}, ed. A.~W. Alsabti \& P.~Murdin (Cham: Springer International Publishing), 1575--1604, \dodoi{10.1007/978-3-319-21846-5_4}

\bibitem[{Johns(2023)}]{Johns:2023a}
Johns, L. 2023, Physical Review Letters, 130, 191001, \dodoi{10.1103/PhysRevLett.130.191001}

\bibitem[{Kato {et~al.}(2024)Kato, Nagakura, \& Johns}]{Kato:2024}
Kato, C., Nagakura, H., \& Johns, L. 2024, Physical Review D, 109, 103009, \dodoi{10.1103/PhysRevD.109.103009}

\bibitem[{Kohyama {et~al.}(1986)Kohyama, Itoh, \& Munakata}]{Kohyama:1986}
Kohyama, Y., Itoh, N., \& Munakata, H. 1986, The Astrophysical Journal, 310, 815, \dodoi{10.1086/164734}

\bibitem[{Langanke \& {Mart{\'i}nez-Pinedo}(2003)}]{Langanke:2003}
Langanke, K., \& {Mart{\'i}nez-Pinedo}, G. 2003, Reviews of Modern Physics, 75, 819, \dodoi{10.1103/RevModPhys.75.819}

\bibitem[{Liu {et~al.}(2023{\natexlab{a}})Liu, Akaho, Ito, Nagakura, Zaizen, \& Yamada}]{Liu:2023c}
Liu, J., Akaho, R., Ito, A., {et~al.} 2023{\natexlab{a}}, Physical Review D, 108, 123024, \dodoi{10.1103/PhysRevD.108.123024}

\bibitem[{Liu {et~al.}(2024)Liu, Nagakura, Akaho, Ito, Zaizen, Furusawa, \& Yamada}]{Liu:2024}
Liu, J., Nagakura, H., Akaho, R., {et~al.} 2024, Muon-Induced Collisional Flavor Instability in Core-Collapse Supernova,  arXiv.
\newblock \doarXiv{2407.10604}

\bibitem[{Liu {et~al.}(2023{\natexlab{b}})Liu, Zaizen, \& Yamada}]{Liu:2023}
Liu, J., Zaizen, M., \& Yamada, S. 2023{\natexlab{b}}, Physical Review D, 107, 123011, \dodoi{10.1103/PhysRevD.107.123011}

\bibitem[{Mayle {et~al.}(1987)Mayle, Wilson, \& Schramm}]{Mayle:1987}
Mayle, R., Wilson, J.~R., \& Schramm, D.~N. 1987, The Astrophysical Journal, 318, 288, \dodoi{10.1086/165367}

\bibitem[{Mezzacappa {et~al.}(2020)Mezzacappa, Endeve, Messer, \& Bruenn}]{Mezzacappa:2020}
Mezzacappa, A., Endeve, E., Messer, O. E.~B., \& Bruenn, S.~W. 2020, Living Reviews in Computational Astrophysics, 6, 4, \dodoi{10.1007/s41115-020-00010-8}

\bibitem[{Morinaga(2022)}]{Morinaga:2022}
Morinaga, T. 2022, Physical Review D, 105, L101301, \dodoi{10.1103/PhysRevD.105.L101301}

\bibitem[{Morinaga \& Yamada(2018)}]{Morinaga:2018}
Morinaga, T., \& Yamada, S. 2018, Physical Review D, 97, 023024, \dodoi{10.1103/PhysRevD.97.023024}

\bibitem[{Nagakura(2023)}]{Nagakura:2023}
Nagakura, H. 2023, Physical Review Letters, 130, 211401, \dodoi{10.1103/PhysRevLett.130.211401}

\bibitem[{Nagakura {et~al.}(2021{\natexlab{a}})Nagakura, Burrows, Johns, \& Fuller}]{Nagakura:2021b}
Nagakura, H., Burrows, A., Johns, L., \& Fuller, G.~M. 2021{\natexlab{a}}, Physical Review D, 104, 083025, \dodoi{10.1103/PhysRevD.104.083025}

\bibitem[{Nagakura {et~al.}(2020)Nagakura, Burrows, Radice, \& Vartanyan}]{Nagakura:2020}
Nagakura, H., Burrows, A., Radice, D., \& Vartanyan, D. 2020, Monthly Notices of the Royal Astronomical Society, 492, 5764, \dodoi{10.1093/mnras/staa261}

\bibitem[{Nagakura {et~al.}(2021{\natexlab{b}})Nagakura, Burrows, \& Vartanyan}]{Nagakura:2021d}
Nagakura, H., Burrows, A., \& Vartanyan, D. 2021{\natexlab{b}}, Monthly Notices of the Royal Astronomical Society, 506, 1462, \dodoi{10.1093/mnras/stab1785}

\bibitem[{Nagakura {et~al.}(2019)Nagakura, Morinaga, Kato, \& Yamada}]{Nagakura:2019}
Nagakura, H., Morinaga, T., Kato, C., \& Yamada, S. 2019, The Astrophysical Journal, 886, 139, \dodoi{10.3847/1538-4357/ab4cf2}

\bibitem[{Nagakura \& Sumiyoshi(2024)}]{Nagakura:2024}
Nagakura, H., \& Sumiyoshi, K. 2024, Neutron Star Kick Driven by Asymmetric Fast-Neutrino Flavor Conversion,  arXiv.
\newblock \doarXiv{2401.15180}

\bibitem[{Nagakura \& Zaizen(2023)}]{Nagakura:2023d}
Nagakura, H., \& Zaizen, M. 2023, Physical Review D, 108, 123003, \dodoi{10.1103/PhysRevD.108.123003}

\bibitem[{Nakazato {et~al.}(2018)Nakazato, Suzuki, \& Togashi}]{Nakazato:2018}
Nakazato, K., Suzuki, H., \& Togashi, H. 2018, Physical Review C, 97, 035804, \dodoi{10.1103/PhysRevC.97.035804}

\bibitem[{Nevins \& Roberts(2024)}]{Nevins:2024}
Nevins, B., \& Roberts, L.~F. 2024, Monthly Notices of the Royal Astronomical Society, 530, 2001, \dodoi{10.1093/mnras/stae1005}

\bibitem[{O'Connor(2015)}]{OConnor:2015}
O'Connor, E. 2015, The Astrophysical Journal Supplement Series, 219, 24, \dodoi{10.1088/0067-0049/219/2/24}

\bibitem[{Pons {et~al.}(1999)Pons, Reddy, Prakash, Lattimer, \& Miralles}]{Pons:1999}
Pons, J.~A., Reddy, S., Prakash, M., Lattimer, J.~M., \& Miralles, J.~A. 1999, The Astrophysical Journal, 513, 780, \dodoi{10.1086/306889}

\bibitem[{Qian \& Woosley(1996)}]{Qian:1996}
Qian, Y.-Z., \& Woosley, S.~E. 1996, The Astrophysical Journal, 471, 331, \dodoi{10.1086/177973}

\bibitem[{Richers(2022)}]{Richers:2022a}
Richers, S. 2022, Physical Review D, 106, 083005, \dodoi{10.1103/PhysRevD.106.083005}

\bibitem[{Richers {et~al.}(2015)Richers, Kasen, O'Connor, Fern{\'a}ndez, \& Ott}]{Richers:2015}
Richers, S., Kasen, D., O'Connor, E., Fern{\'a}ndez, R., \& Ott, C.~D. 2015, The Astrophysical Journal, 813, 38, \dodoi{10.1088/0004-637X/813/1/38}

\bibitem[{Richers \& Sen(2022)}]{Richers:2022b}
Richers, S., \& Sen, M. 2022, in Handbook of {{Nuclear Physics}}, ed. I.~Tanihata, H.~Toki, \& T.~Kajino (Singapore: Springer Nature), 1--17, \dodoi{10.1007/978-981-15-8818-1_125-1}

\bibitem[{Sarikas {et~al.}(2012)Sarikas, {de Sousa Seixas}, \& Raffelt}]{sarikas:2012}
Sarikas, S., {de Sousa Seixas}, D., \& Raffelt, G. 2012, Physical Review D, 86, 125020, \dodoi{10.1103/PhysRevD.86.125020}

\bibitem[{Sawyer(2005)}]{Sawyer:2005}
Sawyer, R.~F. 2005, Physical Review D, 72, 045003, \dodoi{10.1103/PhysRevD.72.045003}

\bibitem[{Sawyer(2016)}]{Sawyer:2016}
---. 2016, Physical Review Letters, 116, 081101, \dodoi{10.1103/PhysRevLett.116.081101}

\bibitem[{Shalgar \& Tamborra(2024)}]{Shalgar:2024}
Shalgar, S., \& Tamborra, I. 2024, Time Dependence of Neutrino Quantum Kinetics in a Core-Collapse Supernova,  arXiv.
\newblock \doarXiv{2406.09504}

\bibitem[{Sigl \& Raffelt(1993)}]{Sigl:1993}
Sigl, G., \& Raffelt, G. 1993, Nuclear Physics B, 406, 423, \dodoi{10.1016/0550-3213(93)90175-O}

\bibitem[{Steiner {et~al.}(2013)Steiner, Hempel, \& Fischer}]{Steiner:2013}
Steiner, A.~W., Hempel, M., \& Fischer, T. 2013, The Astrophysical Journal, 774, 17, \dodoi{10.1088/0004-637X/774/1/17}

\bibitem[{Sugiura {et~al.}(2022)Sugiura, Furusawa, Sumiyoshi, \& Yamada}]{Sugiura:2022}
Sugiura, K., Furusawa, S., Sumiyoshi, K., \& Yamada, S. 2022, Progress of Theoretical and Experimental Physics, 2022, 113E01, \dodoi{10.1093/ptep/ptac118}

\bibitem[{Suzuki \& Nakamura(1993)}]{Suzuki:1993}
Suzuki, Y., \& Nakamura, K. 1993, Frontiers of Neutrino Astrophysics

\bibitem[{Takahashi {et~al.}(1994)Takahashi, Witti, \& Janka}]{Takahashi:1994}
Takahashi, K., Witti, J., \& Janka, H.~T. 1994, Astronomy and Astrophysics, 286, 857

\bibitem[{Tamborra \& Shalgar(2021)}]{Tamborra:2021}
Tamborra, I., \& Shalgar, S. 2021, Annual Review of Nuclear and Particle Science, \dodoi{10.1146/annurev-nucl-102920-050505}

\bibitem[{Togashi {et~al.}(2017)Togashi, Nakazato, Takehara, Yamamuro, Suzuki, \& Takano}]{Togashi:2017}
Togashi, H., Nakazato, K., Takehara, Y., {et~al.} 2017, Nuclear Physics A, 961, 78, \dodoi{10.1016/j.nuclphysa.2017.02.010}

\bibitem[{Volpe(2024)}]{Volpe:2024}
Volpe, M.~C. 2024, Reviews of Modern Physics, 96, 025004, \dodoi{10.1103/RevModPhys.96.025004}

\bibitem[{Wang \& Burrows(2024)}]{Wang:2024b}
Wang, T., \& Burrows, A. 2024, The Astrophysical Journal, 962, 71, \dodoi{10.3847/1538-4357/ad12b8}

\bibitem[{Witt {et~al.}(2021)Witt, Psaltis, Yasin, Horn, Reichert, Kuroda, Obergaulinger, Couch, \& Arcones}]{Witt:2021}
Witt, M., Psaltis, A., Yasin, H., {et~al.} 2021, The Astrophysical Journal, 921, 19, \dodoi{10.3847/1538-4357/ac1a6d}

\bibitem[{Woosley \& Weaver(1995)}]{Woosley:1995}
Woosley, S.~E., \& Weaver, T.~A. 1995, The Astrophysical Journal Supplement Series, 101, 181, \dodoi{10.1086/192237}

\bibitem[{Wu {et~al.}(2015)Wu, Qian, {Mart{\'i}nez-Pinedo}, Fischer, \& Huther}]{Wu:2015}
Wu, M.-R., Qian, Y.-Z., {Mart{\'i}nez-Pinedo}, G., Fischer, T., \& Huther, L. 2015, Physical Review D, 91, 065016, \dodoi{10.1103/PhysRevD.91.065016}

\bibitem[{Xiong {et~al.}(2023{\natexlab{a}})Xiong, Johns, Wu, \& Duan}]{Xiong:2023c}
Xiong, Z., Johns, L., Wu, M.-R., \& Duan, H. 2023{\natexlab{a}}, Physical Review D, 108, 083002, \dodoi{10.1103/PhysRevD.108.083002}

\bibitem[{Xiong {et~al.}(2024)Xiong, Wu, George, Lin, Largani, Fischer, \& {Mart{\'i}nez-Pinedo}}]{Xiong:2024}
Xiong, Z., Wu, M.-R., George, M., {et~al.} 2024, Physical Review D, 109, 123008, \dodoi{10.1103/PhysRevD.109.123008}

\bibitem[{Xiong {et~al.}(2023{\natexlab{b}})Xiong, Wu, {Mart{\'i}nez-Pinedo}, Fischer, George, Lin, \& Johns}]{Xiong:2023}
Xiong, Z., Wu, M.-R., {Mart{\'i}nez-Pinedo}, G., {et~al.} 2023{\natexlab{b}}, Physical Review D, 107, 083016, \dodoi{10.1103/PhysRevD.107.083016}

\bibitem[{Yamada {et~al.}(2024)Yamada, Nagakura, Akaho, Harada, Furusawa, Iwakami, Okawa, Matsufuru, \& Sumiyoshi}]{Yamada:2024}
Yamada, S., Nagakura, H., Akaho, R., {et~al.} 2024, Proceedings of the Japan Academy, Series B, 100, 190, \dodoi{10.2183/pjab.100.015}

\end{thebibliography}
\bibliographystyle{aasjournal}

\end{document}